\documentclass[a4paper,11pt,leqno]{article}

\usepackage{amsmath}
\usepackage{amsfonts}
\usepackage{amsthm}

\usepackage{graphicx}

\newtheorem{theorem}{Theorem}
\newtheorem{prop}{Proposition}

\newtheorem{cor}{Corollary}
\newtheorem{rem}{Remark}

\def\A{{\mathcal A}}
\def\B{{\mathcal B}}
\def\O{{\mathcal O}}

\def\N{\mathbb N}
\def\R{\mathbb R}
\def\Z{\mathbb Z}

\title{Hybrid Quasicrystals, Transport and Localization in Products of Minimal Sets}

\author{T\'ulio O. Carvalho \\ {\it\small Departamento de Matem\'atica -- UEL, CP 6001,
Londrina, PR, 86051-970 Brazil}
\\ and \\
 C\'esar R. de Oliveira\thanks{CRdO thanks the partial support by CNPq.}
\\ {\small\it Departamento de Matem\'atica -- UFSCar, S\~ao Carlos, SP, 13560-970 Brazil}}

\date{\today}

\begin{document}

\maketitle

\begin{abstract} We consider convex combinations of finite-valued almost periodic
sequences (mainly substitution sequences) and put them as potentials of one-dimensional tight-binding  models.  We prove that these sequences are almost periodic.  We call such combinations {\em hybrid quasicrystals} and these studies are related to the minimality, under the shift on both coordinates, of the product space of the respective (minimal) hulls. We observe a rich variety of behaviors on the quantum dynamical transport ranging from localization to transport.
\end{abstract}

\section{Introduction}

The study of transport in one-dimensional aperiodic lattices may be modeled by the
nearest-neighbors tight-binding Hamiltonian (Schr\"odinger operator) in
$l^2(\Z)$
\begin{equation}
\label{hamiltonian} (H\psi)_n=\psi_{n+1}+\psi_{n-1}+\lambda V_n\psi_n,
\end{equation} with~$\lambda>0$ and potentials $V=(V_n)_{n\in\Z}$  generated by aperiodic
sequences. In many circumstances the potentials are real-valued functions of sequences on
a finite set $\A$, called {\em alphabet}; these are models of one-dimensional
quasicrystals \cite{BQ}.

Quantum interferences may lead to localization of the solutions of the corresponding
Schr\"odinger equation
\[ i\frac{\partial}{\partial t}\psi(t)= H\psi(t),
\] as in case of (random) Bernoulli potentials \cite{GdeB}, but also to ballistic motion,
mainly related to  periodic potentials.

Among the characterizations of localization and transport we single out the second moment
of the position operator
\begin{equation}
\label{moments} m_2(T):= \sum_{n=-\infty}^ \infty |n-n_0|^2 |\psi_n(T)|^2,
\end{equation} usually with initial condition concentrated  on a single site $n_0$. For a
large class of potentials the moment
$m_2(T)\le C T^2$ (at least  for  $T>1$) and if $m_2(T)\approx C T^2$ holds we have the
definition of {\em ballistic motion}. Localization will be characterized by a bounded
function $m_2(T)\le C$,
$\forall T$; lack of localization is usually referred to as {\em delocalization} or {\em
transport}. Half the way between these extremes are the {\em anomalous transport}, that is,
\[ m_2(T)\approx C T^\beta \quad\text{with}\quad0<\beta<2,
\] which are usually accompanied by singular continuous spectrum of the operator $H$.
Important examples of such anomalous behavior are the above cited models of
quasicrystals,  among which the most prominent are the (primitive) substitution sequences
\cite{BQ,Q},  for  instance,  Fibonacci, Thue-Morse and Period Doubling sequences. The
Schr\"odinger operators whose potentials are generated by these sequences have singular
continuous spectrum of zero Lebesgue measure (see
\cite{DReview} and references therein).

A widespread spectral point of view makes the association of singular  continuous spectrum
with anomalous transport, absolutely continuous to ballistic motion and  point spectrum of
the Schr\"odinger operator with localization, even though there are known  exceptions,
namely  of operators with purely point spectra showing transport.  Even rank one
perturbation (a very localized one) can exchange point and  singular continuous spectra
\cite{SW}, and the latter surely implies transport  (any continuous spectrum does, as a consequence of RAGE theorem). What about unlocalized perturbations, i.e., those spread over the whole lattice? Certainly  this becomes a too huge class of problems to be reasonably dealt with.

However, there is a special type of such perturbations we think it is worth considering
and may be of some (experimental) relevance in the near future. A particular model of
quasicrystal, as a substitution sequence, is an almost periodic sequence that grows up
from a seed (i.e., an initial condition) and a specific ``growing rule.'' If one has
control of the growing technique, one could grow a quasicrystal in one direction following
one such rule, and in a perpendicular  direction following another rule. This
hybridization creates a potential  which is a linear (convex) combinations of the original
ones. This type of long range perturbations of the potentials can also be considered  from
the theoretical point of view, the sequence spaces are two-dimensional  and have been
considered before
\cite{vuil}.

The potentials we shall consider are constructed as follows. Given two {\em parent}
potentials
$v=(v_n)_{n\in\Z}$,
$u=(u_n)_{n\in\Z}$ and
$0\le
\kappa\le 1$, the {\em hybrid potential} is
\[
\mathcal{I}_\kappa(v,u):=\kappa v+(1-\kappa)u = \left(\kappa
v_n+(1-\kappa)u_n\right)_{n\in\Z}.
\]  Experience with random potentials indicates that if one of them is random then this
characteristic will prevail with respect to localization. If both potentials are periodic,
then the resulting one will also be periodic with period given by their least common
multiple. So, in these extreme cases again, localization and ballistic motion,
respectively, are persistent.  Note that the number of values a hybrid potential assumes
is in general larger than the number of values of each of its components; e.g., if both
$v,u$ take values in
$\{0,1\}$, then  $\mathcal{I}_\kappa(v,u)$ will generally assume all values in
$\{0,\kappa,(1-\kappa),1\}$.

This work is an initial study on this proposal, and we will limit ourselves to almost
periodic potentials taking a finite number of values (notably, substitution sequences). We
present theoretical results on minimality, and data for the moment $m_2(T)$ from
numerical time evolution simulations.

Section~\ref{sectBRSS} review briefly some aspects of finitely valued sequences. In Section~\ref{sectMinimality} we address the question about minimality of the product of minimal sets, giving a sufficient condition for it.  In Section~\ref{sectNumerics} we
report some outcomes of numerical simulations of the moment
$m_2(T)$ for the hybrid potentials, closing in the final section with our concluding
remarks.

\section{Summary on Sequences and Substitutions}\label{sectBRSS}

We denote by $\A^*$ (resp.\ $\A^\Z$) the sets of finite (resp.\ bi-infinite) words with
letters in the finite set
$\A$ (called {\em alphabet}), which can be considered a subset of the real numbers. The
metric on  $\A^\Z$ is
\[ d(a,b) = \begin{cases} 0, \text{ if } \forall n\in \Z, a_n=b_n \\
\frac{1}{2^n}, \text{ where } n=\min\{|j| : a_j\neq b_j\}
            \end{cases} .
 \]   A dynamics on this set is the (left) shift $(\sigma (v))_n=v_{n+1}$. Recall that a
sequence
$v\in\A^\Z$ is almost periodic iff its hull (the bar indicates the closure of the set)
\[
\Omega(v) := \overline{\{\sigma^j(v): j\in\Z\}}
\]is minimal, that is, the hull of any sequence in $\Omega(v)$ coincides with
$\Omega(v)$. The set
$\mathcal{O}(v)=\{\sigma^j(v): j\in\Z\}$ is the orbit of $v$. By Tychonov theorem
$\A^\Z$ is compact and so is every hull as above.

The minimality is an important property of the hull of (primitive) substitution sequences
(see ahead), as well as the existence of a unique ergodic measure, and up to now rigorous
and numerical studies have revealed just one dynamical behavior in each minimal component
(usually rigorous results are restricted to generic or full measure sets). So, as a
first step in the study of such new class of systems, in this work we address the problem of
minimality of the hull of hybrid sequences in case their respective parent potentials are
almost periodic. Setting a product metric, the dynamics with respect to which one considers
minimality, is on the product space of the hulls of the two parent potentials, and is
generated by the (natural) product shift
\begin{equation} \sigma(u,v)_n=(u_{n+1},v_{n+1}) .
\label{prodshift}
\end{equation} We use the same notation for the shift in two and one-dimensional
sequences. In order to investigate the minimality of the product spaces it turns out to be
important  hybridizing not only of $v$ and
$u$, but also  of elements of their one-dimensional orbits; namely, to consider
$\mathcal{I}_\kappa(v,\sigma^j(u))$, for each $j\in\Z$.

A finite word $w$ is indexed
$a_0a_1\cdots a_{|w|-1}$,
$a_i\in \A$, where $|w|$ denotes the length of $w$. Given a set of infinite words $X$, the
{\em language} of
$X$, ${\cal L}(X)$, is the set of finite words occurring in some $w\in X$.

Let us describe some substitution rules which generate sequences of interest for this
work; details can be found in
\cite{BQ,Q}. Given a finite alphabet $\A$ a substitution is a map
$\xi:\A\to \A^* $. Its iterations are defined by concatenation, that is,
$\xi(abc):=\xi(a)\xi(b)\xi(c)$, $\xi^{n+1}(a):=\xi(\xi^n(a))$, $n\ge1$. A substitution is
{\em primitive} if there exists
$k\in\N$ so that for every $a\in\A$ the word $\xi^k(a)$ contains all letters of $\A$. All
substitutions in this work are primitive (see \cite{LdO} for some nonprimitive
substitutions as potentials of Schr\"odinger operators).

A fixed point of a substitution is a sequence $u\in \A^\N$ such that
$\xi(u)=u$. In order to exist, it must be the case that $u_0$ is the first letter of
$\xi(u_0)$. It is known that if
$\xi$ is primitive, there is some $l$ such that $\xi^l$ has a fixed point \cite{Q}, so it
is no loss to assume
$\xi$ has a fixed point.

Fibonacci ({\sc Fcc}), Period Doubling ({\sc PD}) and Thue-Morse ({\sc TM}) substitution
sequences are constructed with an alphabet of two letters
$\{a,b\}$ through the rules
\[
 a\mapsto ab, \;b\mapsto a\; {\rm(Fcc), \;\;\;}\;\;\; a\mapsto ab,
\;b\mapsto ba \;\;\;{\rm(TM),}
\]
\[
 a\mapsto ab, \;b\mapsto aa\; {\rm(PD).}
\] Beginning with $a$ (the seed) and applying successively the substitution rules (the
growing rules), aperiodic sequences are obtained; e.g., the Thue-Morse sequence is given by
\[ abbabaabbaababba\cdots
\]

The paper folding ({\sc PF}) sequence can be obtained with an alphabet of four letters
$\{1,2,3,4\}$, the substitution
\[ 1\mapsto 12, \;\;\;2\mapsto 32,\;\;\; 3\mapsto 14,\;\;\; 4\mapsto 34,
\] (the seed is $1$) and then applying the literal map $1,2\mapsto a$ and
$3,4\mapsto b$.

We then use these substitution sequences to define our potentials $V$; we take
$V_n=-1$ if the $n$-th letter of the sequence is $a$ and $V_n=1$ in case it is $b$. There
are standard ways of extending such substitution potentials for negative values
of~$n$~\cite{BG,HKS}. We do not have to deal with this issue in numerical simulations
because we take a large finite sample of
$N$ sites, using the initial wavefunction concentrated on position
$N/2$, i.e., $\psi_n(t=0)=\delta_{N/2,n}$, $n\ge0$. This is the procedure we  use to
construct almost periodic substitution potentials
$V$.

It is known that the spectrum of the operator~(\ref{hamiltonian}) with finite-valued
aperiodic and almost periodic potentials has no absolutely continuous component 
(primitive substitutions are included) \cite{HKS,K}; although from a rigorous point of
view the lack/presence of eigenvalues in cases of primitive substitution sequences is an open question, as already remarked, no strong evidence of the presence of eigenvalues and localization was found yet.

Given a substitution $\xi$ over a finite alphabet $\A$, denote by $M_\xi$
its substitution matrix,
i.e.,
$M_\xi=a_{w,w'}$, where
$a_{w,w'}$ is the number of occurrences of the letter $w'$ in $\xi(w)$.
$\xi$ is a {\em Pisot substitution} if the dominant eigenvalue of $M_\xi$
has modulus greater than
one, while all the other eigenvalues have absolute values strictly less
than one. For example, the
matrix substitution for {\sc TM} and {\sc Fcc} substitution are
\[ M_{\text{\sc TM}}=\left(
\begin{array}{cc} 1 &   1 \\ 1 &   1\end{array}\right)\quad\text{and}\quad
M_{\text{\sc Fcc}}=\left(
\begin{array}{cc}1 &   1
\\ 1 &   0\end{array}\right),
\] whose dominant eigenvalues are $2$ and $(1+\sqrt{5})/2$, respectively.
The dominant eigenvalue of the {\sc PD} substitution is
$2$, but it is not Pisot,
since the other eigenvalue is $-1$. {\sc PF} is not Pisot either.

\section{Minimality of Hybrid Hulls}\label{sectMinimality}

Let $v$ and $u$ denote almost periodic sequences and $\Omega(v)$,
$\Omega(u)$ be their respective hulls. In the product space
$\Omega(v)\times \Omega(u)$ we have the shift defined by
$\sigma(x,y)=(\sigma(x),\sigma(y))$. This dynamics does not imply that the product space
is minimal if the parent hulls
$\Omega(v)$ and $\Omega(u)$ are minimal. The orbit of a point $(x,y)$ is
$\O(x,y)=\{\sigma^n(x,y) : n\in\Z\}$.  For each $\kappa$, there is a correspondence
between elements of this product space and hybrid sequences
$\mathcal{I}_\kappa(v_l,u_l)$, $v_l\in \Omega(v)$, $u_l\in \Omega(u)$. The potential is a
real-valued function on one such sequence. It is of interest to know whether the potential
is almost-periodic.  We present in this section some results concerning minimality on the
product space of minimal subsets of $\A^\Z$.

Given $\epsilon>0$, the subset of integer numbers
\[ \{n\in\Z :\ d(\sigma^n(x),x)<\epsilon \} \] is called the set of
$\epsilon$-periods of $x\in X\subset \A^\Z$.  When $X$ is minimal, the above set is {\em
syndetic}, i.e., there is an integer
$m$ such that any interval $[n,n+m]\subset \Z$ intersects it. Recall that
$x$ is almost periodic iff that set is syndetic for all
$\epsilon>0$; in this case of finite-valued sequences this is equivalent to the fact that
every finite word in $x$ appears with bounded gaps. 

There is an alternative view of periods in terms of words, or equivalently the cylinder
sets generated by them. If
$a\in\A^\Z$, let
$R_a(w)$ denote the set of integers $n$ such that
\[ a_na_{n+1}\cdots a_{n+|w|-1}=w . \]   Thus $R_a(w)$ is the set of integers $n$ for
which $w$ is a {\em prefix} of
$\sigma^na$. It can be ordered
\[ R_a(w)= \{\alpha_i,i\in\Z:\alpha_i<\alpha_{i+1}\} \]   for some arbitrary choice of
$\alpha_0$. The minimality of
$X$ is equivalent to the fact that for each finite word $w$ that occurs in $X$ there is an
integer $m(w)$ so that 
$\alpha_{k+1}-\alpha_k<m(w)$, $\forall k$ (i.e., $w$ occurs with bounded gaps). Similarly,
for $b\in Y\subset \A^\Z$,
$Y$ minimal, let
$R_b(u)=\{\beta_j,\ j\in\Z:\beta_j<\beta_{j+1}\}$  (the notation should be clear).

In the product space $X\times Y$ we seek a description of the possible minimal sets under
the shift and metric
\[ D((a,b),(c,d)):=d_X(a,c)+d_Y(b,d),\quad a,c\in X,\quad b,d\in Y. \] The existence of
these minimal sets is a consequence of $X\times Y$ compactness and Zorn's Lemma.

Picture $X\times Y$ as the orbit closure of the union of
$(a,\sigma^n(b))$, $n\in \Z$. If we represent the sequence $a$ along a horizontal lattice
$(\cdot,0)\subset \Z^2$ and
$\sigma^r(b)$ along vertical lattices, each attached to the corresponding horizontal
position
$(r,0)$, the orbit  $(a,\sigma^rb)$ is the left translation of the horizontal line
$(\cdot,0)$. Analogously the orbit of
$(\sigma^k(a),b)$ may be followed by pulling horizontally the line at
$(\cdot,k)$. 

We begin to address the question about minimality of $\overline{\O(a,b)}$ by asking if, as
one sits at different positions along the horizontal axis, one sees the same pair of
finite words $u$, $w$ upwards and to the right respectively, infinitely often. While this
certainly happens at each
$\beta_n$ and $\alpha_n$ alternatively upwards and to the right, one is interested in
these words appearing at the same time {\em and} with bounded gaps.

\begin{prop}  If $X\subset \A^\Z$ and $Y\subset \B^\Z$ are minimal sets, then
$X\times Y$ decomposes into finitely many minimal sets.
\label{teo1}
\end{prop}

\begin{proof} Suppose on the contrary that we had infinitely many invariant sets
$M_i$.  Choose a point in each $M_i$ and an open set $U_i$ containing it but with no
intersection with $M_j$, $j>i$. Complete this cover with
$V_i=M_i\setminus U_i$ (recall that $U_i$ is also closed since we are dealing with product
of cylinders). From the cover of $U_i$ and $V_i$'s we cannot extract a finite subcover,
but $X \times Y$ is compact.
\end{proof}

Theorem 1.17 in \cite{furst} yields a point whose orbit closure is a minimal set. We can
show that this holds for every point in $X\times Y$, whenever $X$ and $Y$ are minimal sets.

\begin{prop}\label{propOrbMin}  Suppose $X,Y\subset \A^\Z$ are minimal. Given a point
$z\in X\times Y$, its orbit closure $\overline{\O(z)}$ is minimal.
\end{prop}
\begin{proof}  Pick a point $(a,\sigma^jb)$ from a minimal set $M\subset X\times Y$. Then,
for any
$\epsilon>0$, there exists a sequence $(n_p)_{p\in \N}$, $n_p \nearrow
\infty$, $|n_{p+1}-n_p|$ bounded such that 
\[ D((a,\sigma^j(b)),\sigma^{n_p}(a,\sigma^j(b)))< \epsilon  . \] Now pick a  point
$(\sigma^ka,\sigma^lb)\in X\times Y$. For any
$n_p>h=\max\{|l-j|,|k|\}$ 
\[\begin{split} & D(\sigma^k(a,\sigma^{l-k}(b)),\sigma^{n_p+k}(a,\sigma^{l-k}(b)))= \\ &\;
d_X(\sigma^k a,\sigma^{n_p+k}a)+d_Y(\sigma^lb,\sigma^{n_p+l}b) \\ &\; \le 2^k
d_X(a,\sigma^{n_p}a)+ 2^{|l-j|}d_Y(\sigma^j b,\sigma^{n_p+j}b) <2^{h+1}\epsilon
\end{split} \] and this can be made arbitrarily small. Since any $z\in X\times Y$ belongs
to the closure of the orbit of some $(\sigma^ka,\sigma^lb)$, the proposition is proved.
\end{proof}

This proves the assertion in the abstract
\begin{cor}  If $X,Y\subset \A^\Z$ are minimal sets, then a sequence $z\in X\times Y$ is
almost periodic, as well as any sequence obtained from it by some real-valued function
defined on $X\times Y$.
\end{cor}

In what follows, unless stated on the contrary, we assume that $X$ and $Y$ are minimal
sets. Now the question is to characterize when the product
$X\times Y$ is minimal.

\begin{prop}  Suppose there exists a sequence $n_k\nearrow \infty$ so that
$\sigma^{n_k}a\rightarrow a^*$ and $\{\sigma^{n_k+l}b: n_k\}$, for some $l$ fixed, is dense
in
$Y$. Then $X\times Y$ is minimal.
\end{prop}
\begin{proof} Our hypothesis asserts that $(a^*,Y)$ is contained in the orbit closure
$\overline{\O(a,\sigma^l(b))}$. Due to Proposition~\ref{propOrbMin}, it is enough to show
that there is a dense orbit in $X\times Y$. Given
$(x,y)\in X\times Y$ and $\epsilon>0$
\[ \begin{split}  &D((\sigma^na,\sigma^{n+l}b),(x,y)) \\  &\ \leq
D((\sigma^na,\sigma^{n+l}(b)),(\sigma^j(a^*),y))+D((\sigma^j(a^*),y),(x,y))
\ .
\end{split}
\] Since $X$ is minimal, the second term may be made less than $\epsilon/3$, and this
fixes $j$. Now 
\[ D((\sigma^na,\sigma^{n+l}(b)),(\sigma^j(a^*),y))
=d(\sigma^n(a),\sigma^j(a^*))+d(\sigma^{n+l}(b),y)  . \] We note that along the
subsequence $n=n_k+j$ the first term is less than $\epsilon/3$ for every $n_k$
sufficiently large. The set of points $\sigma^{n_k+l}b$, $n_k>N$, is dense in $Y$, $N$ a
fixed arbitrary integer. Therefore, if
$z=\sigma^{-j}y$, there exists $n_k$ so that
$d(\sigma^{n_k+l}b,z)<{\epsilon'}$. Choosing $\epsilon'$ small enough yields
$d(\sigma^{n_k+(j+l)}b,y)<\epsilon/3$, for some $n_k>N$.
\end{proof}

Therefore, if $X\times Y$ is not minimal, then for every convergent sequence
$\sigma^{n_k}(a)\rightarrow a^*$ one has that $\sigma^{n_k+l}(b)$ is not dense, for any
$l$.

We know that the dynamics of a map on the torus ${\mathbb T}^d$:
$T({\mathbf\theta})={\mathbf\theta}+{\mathbf\alpha}\pmod 1$ is ergodic  when ${\mathbf
\alpha}$ is rationally independent. By coding this dynamics with a partition along each
circle, we get a symbolic sequence which is semi-conjugate to the original dynamics
\cite{alessandri}. This is an example of a product of two sequences spaces which is
minimal.  We say that $(w,u)$ is a {\em prefix} of $(a,b)$ if $w$ is a prefix of $a$ and
$u$ is a prefix of
$b$. It is easy to characterize lack of minimality in terms of the language of $X\times
Y$. Indeed, if ${\cal L}(a,b)$ denote the set of words of $(a,b)\in X\times Y$, we have
${\cal L}(a,b)\subset {\cal L}(a)\times {\cal L}(b)$. Hence, if $X\times Y$ is not
minimal, for each invariant set $M\subset X\times Y$, there are words $r\in {\cal L}(a)$
and
$s\in {\cal L}(b)$ such that $(r,s)$ does not occur in $M$. 
\begin{rem}  If $a$ is an almost periodic sequence, note that ${\cal L}(a)={\cal
L}(\Omega(a))$.
\end{rem}

By hull of a substitution we understand the hull of any of its fixed points. For primitive
substitution sequences we get a simple criterion for minimality of the product of their
hulls. Recall that this case is our choice of prototypes of hybrid quasicrystals. The
argument comes from the proof of a result Hansel in
\cite{han} related to Cobham's Theorem (see also \cite{cob,dur}). Recall that two positive
numbers $\theta$ and
$\vartheta$ are multiplicatively independent if the equation $\theta^l=\vartheta^k$ holds
only for $l=k=0$.

\begin{theorem}\label{thmMultIndepMin}  Let $\xi$ and $\zeta$ be two primitive
substitutions on the (finite) alphabets
$\A$ and $\B$, respectively, and denote by $X$ and $Y$  their respective hulls under the
shift.  If $M_\xi$ and
$M_\zeta$ have multiplicatively independent dominant eigenvalues, then $X\times Y$ is
minimal.
\end{theorem}

\begin{proof}   Let $a=\xi(a)=(a_j)_{j\in\Z}$ and $b=\zeta(b)=(b_j)_{j\in\Z}$ be fixed
points of the corresponding substitutions and $M=\overline{\O(a,b)}$. If $X\times Y$ is
not minimal, then $M\ne X\times Y$ and there is a finite word $(r,s)$ in $X\times Y$ that
does not occur in $M$. Thus, for any $r_0,r_1,s_0,s_1$,
$|r_0|=|s_0|$, $(r_0rr_1,s_0ss_1)$ does not occur in the orbit of $(a,b)$ either.

Since the substitutions are primitive, there is a $k$ so that for all $n\geq k$ the words
$\xi^n(w)$, $w\in\A$, contain
$r$, and $\zeta^n(u)$, $u\in \B$, contain $s$. We choose $r_0$ and $r_1$ so that
$\xi^n(w)=r_0rr_1$ above. Then $s_0$ and $s_1$ are chosen so that $\zeta^n(u)$ contains
$s_0ss_1$. We conclude that for any pair $(w,u)\in \A\times\B$ there is some $n_0$ (which
may be taken {\em big}) so that 
\begin{equation} (\xi^{n_0}(w),\zeta^{n_0}(u)) \text{ is not a prefix of }
\sigma^k(a,\sigma^{-l}b) \text{ for every } k \text{ and some } l.
\label{one}
\end{equation}

Let $\theta$ and $\vartheta$ be the dominant eigenvalues of $M_\xi$ and
$M_\zeta$ respectively. Consider the subsets of $\N$
\begin{align*}  & E(X)=\{|\xi(a_0a_1\cdots a_m)|, m>0\}, \\   & E(Y)=\{|\zeta(b_0b_1\cdots
b_m)|, m>0\}  .
\end{align*}
$E(X)$ contains some of the positions where the words $\xi^j(w),\ \forall j, \forall
w\in\A$, occur in $a$. By Lemma~2 in \cite{han}, for large enough $m$, these positions are
the integer numbers closest to $a\theta^{pj}+b$ for some integer $p>0$, and real $a>0,b$.
The same holds for $E(Y)$, in that it contains integers closest to
$a'\vartheta^{qj}+b'$, for some integer $q>0$, and real $a'>0,b'$. But $\theta^p$ and
$\vartheta^q$ are multiplicatively independent, so the set of ratios
\[\frac{a\theta^{pj}+b}{a'\vartheta^{qj}+b'}
\] is dense in $\R^+$. Therefore, it must be
the case that the intersections
$E(X)\cap E(Y)$ and $E(X)\cap \{E(Y)-l\}$, for any $l$, are not empty. This contradicts
(\ref{one}).
\end{proof}

\begin{rem}
In the case of Pisot substitutions, the proof is simpler in that the $E(X)$ will contain integer numbers close to $\theta^{pj}$, while $E(Y)$ contains integers close to $\vartheta^{qj}$, for integer $j>0$. If $\theta$ and $\vartheta$ are multiplicatively independent, the same argument follows.
\end{rem}

\begin{cor} \label{corolExemMin} If $\Omega_{\text{\sc TM}}$, $\Omega_{\text{\sc Fcc}}$ and
$\Omega_{\text{\sc PD}}$ are the hulls of the indicated substitution sequences, then the
products $\Omega_{\text{\sc TM}}\times\Omega_{\text{\sc Fcc}}$ and $\Omega_{\text{\sc
PD}}\times\Omega_{\text{\sc Fcc}}$ are minimal.
\end{cor}

One can investigate whether some product spaces generated by constant length substitutions
are not minimal in a case by case analysis. For instance if
$\xi$ denotes the Thue-Morse substitution and $\eta$ is the period-doubling substitution,
then contained in the above described $X\times Y$, one has the following subset $\Omega$.
Let $\B$ denote the four letter alphabet:
$\{(a,a),(a,b),(b,a),(b,b)\}$. On $\cal B$ we define the substitution
\[ \zeta(x,y):=(\xi(x),\eta(y)) . \] Explicitly, $\zeta(a,a)=(ab,ab)=(a,a)(b,b)$,
$\zeta(a,b)=(ab,aa)=(a,a)(b,a)$,
$\zeta(b,a)=(ba,ab)=(b,a)(a,b)$ and $\zeta(b,b)=(ba,aa)=(b,a)(a,a)$. This substitution
$\zeta$ can be shown to be primitive with two fixed points. The fixed points  belong to
the same hull, since the languages of the fixed points of a primitive substitution
coincide. In a four letter alphabet, Berstel has considered a substitution isomorphic to
$\zeta$ when constructing square-free words
\cite{bers}. Let $u=abbabaab\cdots$ be one fixed point of the Thue-Morse substitution and
$w=abaaabab\cdots$ be the aperiodic fixed point of the period-doubling substitution. We
can see that $\Omega$ is an invariant minimal subset strictly contained in $X\times Y$ by
noticing that, while $(abba,baaa)$ is a prefix of $(u,\sigma w)$, it does not occur in any
point of the orbit $\sigma^j(u,w)$. 

Similarly, we have analyzed the substitutions defined in an eight
letter alphabet by the product of period doubling and Rudin-Shapiro,
and the product of period doubling and paper-folding. These
substitutions are {\em semi-primitive}, in the sense of \cite{BG}, see
also \cite{ltww} where semi-primitiveness is shown for Rudin-Shapiro
substitution. There is a sub-alphabet, with six letters, in which they
are primitive. These substitutions also have two fixed points. The
same argument on the location of the letter $b$ in the period doubling
substitution leads to more than one invariant set in $X\times Y$. 

\section{Numerics of the Moment}\label{sectNumerics}

In this section we report some numerical simulations of the moment
$m_2(T)$ as a function of time $T$ for some hybrid quasicrystals. Basis sets were usually
of size $2^{14}$, and the time evolution was done by integrating the Schr\"odinger
equation using a sympletic integrator, as described in~\cite{cestu}.  The emphasis will be
on hybrid substitution quasicrystals. It is expected that different minimal sets present
different behavior of $m_2(T)$ and, with respect to numerics, this is the working setting
accepted here. 

In these numerical experiments we have mostly fixed
$\kappa=1/2$, but exceptions are explicitly mentioned. We also set $\lambda=1$ 
(preliminary results indicate that the qualitative behavior is independent of
$\lambda\ne0$). The guide to the simulations was based on two properties used in
Theorem~\ref{thmMultIndepMin}, that is, the  multiplicatively independent dominant
eigenvalues of their substitution matrices. 
\begin{figure}[ht]
\begin{center}
\includegraphics [width=4.0in,height=2.5in] {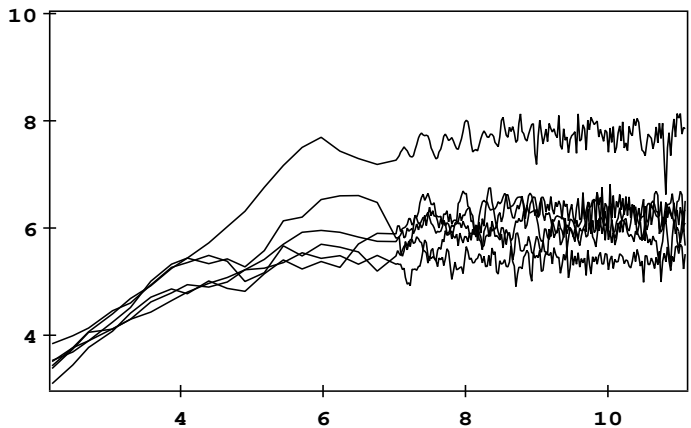}
\caption{The moment as function of time ($\log-\log$ scale) for the combination of {\sc
TM} and {\sc Fcc} substitution sequences. The sequence {\sc Fcc} was kept fixed, while
{\sc TM} was shifted by $0,1,\cdots,5$ in order to explore different elements of the
product of their hulls.
\label{figTMFcc}}
\end{center}
\end{figure}

\begin{figure}[ht]
\begin{center}
\includegraphics [width=4.0in,height=2.5in] {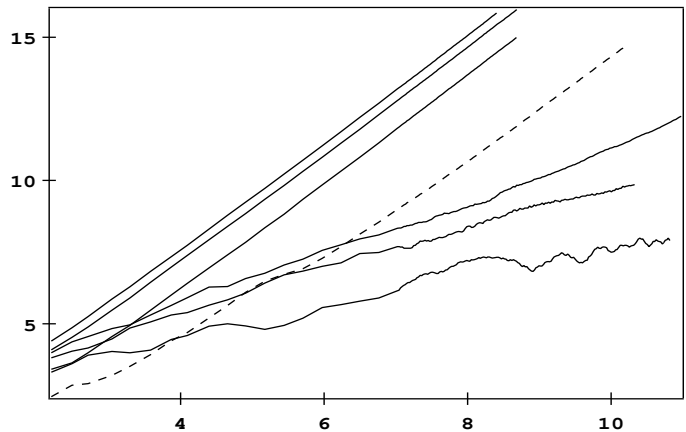}
\caption{The moment as function of time ($\log-\log$ scale) for the combination of {\sc
TM} with itself shifted. The dashed line is for the original {\sc TM} (no shift at all).
\label{figTMTM}}
\end{center}
\end{figure}

First consider the hybridizing of {\sc TM} and {\sc Fcc}. The results are summarized in
Figure~\ref{figTMFcc}.   Different elements of the product of the hulls are obtained by
keeping one sequence fixed and shifting the other before the combination. Although both
sequences individually generate transport (for {\sc TM} see the dashed line in
Figure~\ref{figTMTM}), when combined we have got only one behavior, in accordance with
Theorem~\ref{thmMultIndepMin} and Corollary~\ref{corolExemMin}, since the hybrid  hull is
minimal in this case. This gives an example of numerical dynamical localization in an
almost periodic sequence. As a complement to such simulations we have also considered
$\kappa=0.2$ and $0.8$, and localization was always found; again the minimality seems to
be the important property. 

Another possibility we have investigated is when the two
involved substitutions have multiplicative dependent dominant eigenvalues. The extreme
case is for equal eigenvalues and we have selected this situation by hybridizing a
substitution with shifts of itself. Figure~\ref{figTMTM} presents the results of these
simulations for {\sc TM} sequence; transport was found in  all cases, although with
different exponents $\beta$, indicating the presence of more than one minimal component in
the product of the hulls; so  there are different hybrid quasicrystals in this case. In
Figure~\ref{figTMTM}  the dashed line is for the original {\sc TM} sequence. We have noted
three distinct behaviors, with the dashed line as a border between them: for some shift
values the moment follows the dashed line ($\beta\approx \beta_{\text{\sc TM}}=1.8$),
others present a range of $0<\beta< \beta_{\text{\sc TM}}$ values, while others with near
ballistic behavior (i.e., $\beta> 1.9$); that is, if $\beta>\beta_{\text{\sc TM}}$ then it
is near the maximum possible value. We add that for the combination of {\sc Fcc} with
itself  similar results were obtained (not shown), that is, transport prevails and
different exponent values of $\beta$ were found; however, without a case near the
ballistic motion. 

\begin{figure}[ht]
\begin{center}
\includegraphics [width=4.0in,height=2.5in] {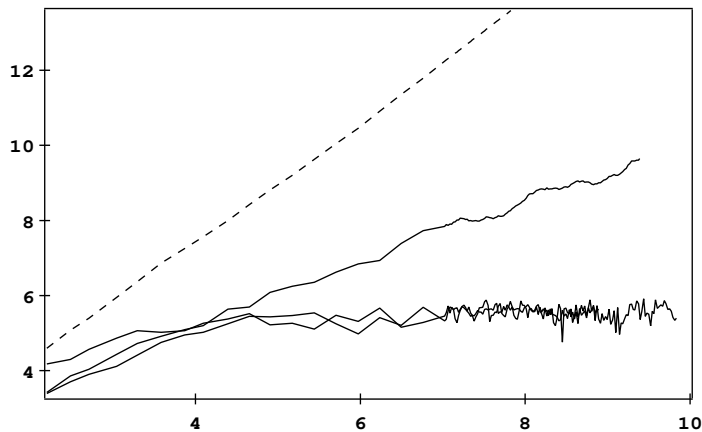}
\caption{The moment as function of time ($\log-\log$ scale) for the combination of {\sc
PF} with periodic sequences. The periods were 4 (dashed), 16 (line), 7 and 10
(localization).\label{figpfper}}
\end{center}
\end{figure}

The same procedure was applied to the {\sc PD} substitution. If no shift is applied to the
sequences, then the original sequence is obtained and it cannot be considered a hybrid
quasicrystal,  although it is embedded in the product space. Except for this case, where 
$\beta_{\text{\sc PD}}\approx 1.78$, all other simulations clearly indicate a motion near
the ballistic one (no figure is shown). It appears that the self-product of
period-doubling substitution contains only two minimal components.  We have also combined
almost periodic substitution sequences with periodic ones (with periods up to 32), and
quite distinct behaviors were found. A periodic sequence is also almost periodic and its
hull has finitely many elements (as many as its period).  We have hybridized {\sc PD},
{\sc TM}, {\sc PF} and {\sc Fcc} with periodic sequences and, depending on the choice of
the period, for some cases we have found transport, with different values of $\beta$, but
in some other periods we got localization. Figure~\ref{figpfper} shows some instances of
{\sc PF} combined with periodic sequences. Such long range perturbations have shown a rich
range of possibilities.

\section{Conclusions}

In this work we considered hybrid quasicrystals, defined by the convex combination two
parent finitely valued almost periodic sequences, as new models of one-dimensional
quasicrystals. Hybridization of substitution sequences was given special attention.

We investigated in some generality the minimality of product spaces $X\times Y$, when both
$X$ and $Y$ are minimal, and Section~\ref{sectMinimality} presented a sufficient condition
for primitive substitutions, which is the multiplicative independence of the eigenvalues
of their substitution matrices. Minimality is well known when the metric on the
sequence space is given by the sup-norm \cite{pete}, but requires extra work in the
setting of finitely valued sequences.

Some hybrid  potentials were inserted into Schr\"odinger equation and the time evolution
of concentrated initial conditions numerically investigated; the interest was in
localization and transport in such structures. In order to classify our numerical results
we have adopted the pragmatic position that elements in the same minimal set should
generate similar time evolutions. This was confirmed in  cases our analytical results
proved minimality for the product of minimal hulls, and suggested the presence of more
than one minimal component in other cases. The figures presented in
Section~\ref{sectNumerics} illustrate these behaviors. The hybridization with periodic
sequences was also numerically considered.

The numerical results suggest a rigorous investigation of localization in some hybrid
quasicrystals. This could be accomplished by proving that the Lyapounov exponent $\gamma$
in these sequences is uniformly positive, that is, the existence of $c>0$ such that
$\gamma>c>0$.

To our knowledge, this result would be relevant since minimal sequences generated by
primitive substitutions have been shown to have zero Lyapounov exponent by the following
reasoning. Recall that ${\cal L}(\Omega)$ denotes the language of the
minimal subshift $\Omega$, and let $[v]$ be the cylinder set defined by the word $v$:
\[ [v]\equiv \{\omega\in \Omega: \omega_1\cdots \omega_{|v|}=v\} \] Let $\nu$ be a
$\sigma$-invariant probability on
$(\Omega,\sigma)$ and $n\in \N$ and ${\cal L}_n(\Omega)$ the set of words of length $n$
occurring in $\Omega$. Define 
\[ \eta_\nu(n)=\min\{\nu([w]):w\in L_n(\Omega)\} \] Boshernitzan's condition, first
considered in subshifts related to interval exchange transformations \cite{bosh}, may be
written as
\[ \limsup_{n\rightarrow \infty} n\eta_\nu(n)>0 \ . \] It is proven that the family of
ergodic operators
$(H_\omega)_{\omega\in \Omega}$, when $\Omega$ is a minimal subshift satisfying
Boshernitzan's condition, have zero Lyapounov exponent everywhere in their spectrum, which
is a Cantor set of zero Lebesgue measure \cite{dalebos}.

We can see that Boshernitzan's condition does not hold in hybrid quasicrystals. For an
almost periodic hybrid sequence
$z\in \Omega$, the complexity $p_z(n)$, which counts the number of words of length $n$ in
$z$, is at least of the order of $n^2$, because the complexity of each component in a
hybrid sequence is at least of order $n$. On the other hand, the measure of any cylinder
must be inversely proportional to the complexity, because  
\[ p_z(n)\min_{v\in {\cal L}_n(\Omega)} \nu([v])< \sum_{v\in {\cal L}_n(\Omega)} \nu([v])
= 1 \ .\] for any probability
$\nu$.

This leaves one important theoretical question, to pursue the possibility of Anderson
localization in minimal hybrid quasicrystals.

\end{document}